\documentclass[aps,prb,twocolumn,superscriptaddress,showpacs]{revtex4-1}

\usepackage{graphicx}
\usepackage{amsmath}
\usepackage{bm}
\usepackage{longtable}
\usepackage{xspace}

\newcommand{\rc}{$\alpha$-RuCl$_3$\xspace}
\newcommand{\nio}{Na$_2$IrO$_3$\xspace}
\newcommand{\lio}{Li$_2$IrO$_3$\xspace}

\begin{document}

\title{The monoclinic crystal structure of $\alpha$-RuCl$_3$ and \\
the zigzag antiferromagnetic ground state}

\author{R.\,D. Johnson}
\email{roger.johnson@physics.ox.ac.uk} \affiliation{Clarendon
Laboratory, University of Oxford Physics Department, Parks Road,
Oxford, OX1 3PU, United Kingdom}
\affiliation{ISIS Facility,
Rutherford Appleton Laboratory-STFC, Chilton, Didcot, OX11 0QX,
United Kingdom}
\author{S. C. Williams}
\affiliation{Clarendon Laboratory, University of Oxford Physics
Department, Parks Road, Oxford, OX1 3PU, United Kingdom}
\author{A.\,A. Haghighirad}
\affiliation{Clarendon Laboratory, University of Oxford Physics
Department, Parks Road, Oxford, OX1 3PU, United Kingdom}
\author{J. Singleton}
\author{V. Zapf}
\affiliation{National High Magnetic Field Laboratory MPA-NHMFL,
TA-35, MS-E536 Los Alamos National Laboratory, Los Alamos, NM
87545, USA}
\author{P. Manuel}
\affiliation{ISIS Facility,
Rutherford Appleton Laboratory-STFC, Chilton, Didcot, OX11 0QX,
United Kingdom}
\author{I.\,I. Mazin}
\affiliation{Code 6393, Naval Research Laboratory, Washington, DC 20375, USA}
\author{Y. Li}
\author{H.\,O. Jeschke}
\author{R. Valent{\'\i}}
\affiliation{Institut f\"{u}r Theoretische Physik,
Goethe-Universit\"{a}t Frankfurt, 60438 Frankfurt am Main,
Germany}
\author{R. Coldea}
\affiliation{Clarendon Laboratory, University of Oxford Physics
Department, Parks Road, Oxford, OX1 3PU, United Kingdom}
\date{\today}
\begin{abstract}
The layered honeycomb magnet \rc has been proposed as a candidate
to realize a Kitaev spin model with strongly frustrated,
bond-dependent, anisotropic interactions between spin-orbit
entangled $j_\mathrm{eff}=1/2$ Ru$^{3+}$ magnetic moments. Here we
report a detailed study of the three-dimensional crystal structure
using x-ray diffraction on un-twinned crystals combined with
structural relaxation calculations. We consider several models for
the stacking of honeycomb layers and find evidence for a parent
crystal structure with a monoclinic unit cell corresponding to a
stacking of layers with a unidirectional in-plane offset, with
occasional in-plane sliding stacking faults, in contrast with the
currently-assumed trigonal 3-layer stacking periodicity. We report
electronic band structure calculations for the monoclinic
structure, which find support for the applicability of the
$j_\mathrm{eff}=1/2$ picture once spin orbit coupling and electron
correlations are included. Of the three nearest neighbour Ru-Ru
bonds that comprise the honeycomb lattice, the monoclinic
structure makes the bond parallel to the $b$-axis non-equivalent
to the other two, and we propose that the resulting differences in
the magnitude of the anisotropic exchange along these bonds could
provide a natural mechanism to explain the previously reported spin gap in powder inelastic neutron scattering measurements, in contrast to spin models based on the three-fold symmetric trigonal structure, which predict a gapless spectrum within linear spin wave theory. Our susceptibility measurements on both powders and
stacked crystals, as well as magnetic neutron powder diffraction
show a single magnetic transition upon cooling below
$T_\mathrm{N}\approx$13~K. The analysis of our neutron powder
diffraction data provides evidence for zigzag magnetic order in
the honeycomb layers with an antiferromagnetic stacking between
layers. Magnetization measurements on stacked single crystals in
pulsed field up to 60~T show a single transition around 8~T for
in-plane fields followed by a gradual, asymptotic approach to
magnetization saturation, as characteristic of
strongly-anisotropic exchange interactions.
\end{abstract}

\maketitle

\section{Introduction}

There has been considerable recent interest in materials that
realize strongly-anisotropic, bond-dependent interactions as the
resulting frustration effects could potentially stabilize novel
forms of cooperative magnetic order or a spin liquid
state.\cite{review_iridates} A canonical Hamiltonian is the Kitaev
spin model on the honeycomb lattice,\cite{Kitaev2006} where each
bond carries an Ising interaction, but where the Ising axes are
reciprocally orthogonal for the three bonds meeting at each
lattice site, leading to an exotic quantum spin liquid state with
fractional spin excitations. In a pioneering set of
papers\cite{jackeli,chaloupka} it was proposed that Kitaev physics
may be realized in ${\cal A}_2$IrO$_3$ (${\cal A}=$ Na, Li)
materials with a tri-coordinated, edge-sharing bonding geometry of
IrO$_6$ octahedra. Here the combined effect of strong spin-orbit
coupling at the Ir$^{4+}$ $5d^5$ site and near-cubic crystal field
of the O$_6$ octahedra stabilize $j_\mathrm{eff}=1/2$ Ir moments,
and superexchange via two near 90$^\circ$ Ir-O-Ir paths is
predicted to couple (to leading order) only the magnetic moment
components normal to the plane of the Ir-O-Ir bond, with three
such near-orthogonal planes meeting at each Ir site. Evidence for
dominant Kitaev interactions in such materials has been observed
in the structural polytypes $\beta$- and $\gamma$-Li$_2$IrO$_3$
where the Ir ions have the same local threefold coordination as in
the planar honeycomb, but now form fully-connected
three-dimensional networks, so-called hyperhoneycomb and
stripyhoneycomb, respectively. In both structural polytypes
complex counter-rotating and non-coplanar incommensurate magnetic
orders have been observed,\cite{BiffinPRL, BiffinPRB} which cannot
be reproduced by isotropic (Heisenberg) exchanges, but require the
presence of dominant ferromagnetic Kitaev
interactions\cite{Itamar,Lee1,Lee2} supplemented by additional
smaller interactions. In contrast, the layered honeycomb iridate
Na$_2$IrO$_3$ shows a very different magnetic order, with spins
arranged in zigzag ferromagnetic chains aligned
antiferromagneticaly,\cite{Hill,choi12,ye} believed to be
stabilized by the competition between many interactions including
a strong ferromagnetic Kitaev term and further neighbor
interactions.\cite{Khaliullin} In Na$_2$IrO$_3$ evidence for the
presence of strong Kitaev interactions has been provided by
measurements of the diffuse scattering at temperatures above the
magnetic ordering transition temperature, which observed a locking
of the polarization of spin fluctuations with the wavevector
direction.\cite{Chun15}

\rc has been proposed\cite{Plumb14} as a candidate Kitaev material
in a $4d$ analogue of the layered honeycomb iridates. This might
be surprising at first as the spin-orbit coupling is expected to
be considerably weaker in Ru compared to Ir (due to the smaller
atomic number), but it was argued\cite{Plumb14,Heung15} that i) the
crystal field of the Cl$_6$ octahedra may potentially be much
closer to cubic in \rc as layers are only very weakly bonded (by
van der Waals interactions), in contrast to Na$_2$IrO$_3$ where
the O$_6$ octahedra are strongly trigonally squashed due to the
strong bonding to the adjacent hexagonal Na$^+$ layers, and ii)
correlation effects in a narrow band could potentially enhance the
effects of spin-orbit coupling.

The magnetic properties of \rc are currently the subject of much
experimental and theoretical
investigation.\cite{Plumb14,sears15,Banerjee15,kubota15,majumder15,SandilandsMar15,Sandilands15,Rousochatzakis15}
Early studies have established the existence of two distinct
structural polytypes: the $\alpha$ polytype with edge-sharing
RuCl$_6$ octahedra forming stacked honeycomb layers with magnetic
order below $\approx$14~K [Ref.~\onlinecite{Fletcher67}], and the
$\beta$ polytype with face-sharing RuCl$_6$ octahedra arranged in
chains, which shows no magnetic ordering down to the lowest
temperatures measured.\cite{kobayashi92} However, detailed studies
of the three-dimensional crystal structure of the layered
($\alpha$) polytype have proved difficult because of the
prevalence of diffuse scattering due to stacking
faults,\cite{Brodersen65} an inevitable consequence of the weak
bonding between adjacent honeycomb layers. A trigonal space group
$P\,3_112$ with a 3-layer stacking periodicity is usually
presupposed based on an early structural study,\cite{stroganov57}
although this structural model has been questioned by later
studies.\cite{vonSchnering1966,Brodersen68,Cantow1990} In
particular, Ref.~\onlinecite{Brodersen68} reported a monoclinic
$C2/m$ stacking of honeycomb layers for the related halide
IrBr$_3$ (AlCl$_3$ structure type\cite{Ketelaar}) and proposed, by
analogy, a similar structural framework for \rc, but no lattice
parameters or any other structural details were provided. The
difficulty in reliably solving the crystal structure stems from
the fact that in principle several candidate stacking sequences of
the honeycomb layers may be possible (monoclinic, trigonal,
rhombohedral - to be discussed later) and it is experimentally
rather challenging to reliably distinguish between them in the
presence of stacking faults and/or when samples may contain
multiple twins. Having a reliable determination of the full
three-dimensional crystal structure is important for understanding
the underlying electronic and magnetic properties, as electron
hopping terms, and consequently magnetic interactions and
anisotropies, appear to be quite sensitive to the stacking
sequence of layers and to weak distortions inside each layer, as
we will show later in Sec.~\ref{sec:electronic}.

Previous studies on single crystals of \rc have observed two
anomalies near 8 and 14~K in both magnetic susceptibility and heat
capacity \cite{sears15,majumder15,kubota15} (with the transition
near 8~K attributed\cite{sears15} to the onset of zigzag magnetic
order as in \nio), whereas studies on powder samples showed only
one anomaly near $T_\mathrm{N}\approx$13~K
[Refs.~\onlinecite{Fletcher67,kobayashi92}], raising the question
of why the powders and single crystals show distinct behaviors. To
date, the ground state magnetic structure is yet to be reported
for samples that exhibit a single magnetic phase transition upon
lowering temperature.

Here, we report comprehensive results and an extensive discussion
of x-ray diffraction measurements on {\em un-twinned} crystals of
\rc that display a single magnetic phase transition upon cooling
to low temperatures, in agreement with powder samples. We find
that the crystal structure is monoclinic, with space group $C2/m$.
Features in the diffraction pattern necessitated by the assumed
trigonal $P\,3_112$ model are clearly absent. The monoclinic
structure of \rc is found to be iso-structural to the layered
honeycomb materials \nio [Ref.~\onlinecite{choi12}] and
$\alpha$-\lio [Ref.~\onlinecite{Malley08}]. From neutron powder
diffraction data, we present evidence of a magnetic propagation
vector, \textbf{k}=(0,1,0.5), consistent with \emph{zigzag} or
\emph{stripy} long-range magnetic ordering. We find that the
calculated magnetic diffraction pattern expected for the stripy
model is inconsistent with the experimental data and conclude that
the zigzag model with antiferromagnetic stacking gives the best
account of the true magnetic structure. Furthermore, we
characterize the stability of the zigzag order in applied magnetic
field and construct a magnetic phase diagram for field applied in
the honeycomb layers. To complement the x-ray diffraction studies
we report electronic band structure calculations to check the
stability of the crystal structure and determine the resulting
magnetic ground state of the Ru$^{3+}$ ions.

The paper is organized as follows: Sec.~\ref{sec:methods} presents
the methods employed. Single crystal diffraction
results are given in Sec.~\ref{sec:xrd}, with the space-group
determination and stacking faults analysis presented in
Sec.~\ref{sg}, the structural refinement discussed in
Sec.~\ref{sr}, and comparison to other structural models drawn in
Sec.~\ref{sc}. Following this, in Sec.~\ref{sec:magnetism} we
focus on the magnetic order at low temperatures through discussion
of susceptibility, pulsed-field magnetization and neutron powder
diffraction results. In Sec.~\ref{ins} we discuss the
implications of the monoclinic symmetry for the low-energy spin
excitations and in Sec.~\ref{sec:electronic} we present results
of {\em ab-initio} electronic structure calculations. Finally,
conclusions are summarized in Sec.~\ref{sec:con}.

\section{\label{sec:methods}Methods}
Crystals of \rc were grown by vacuum sublimation from commercial
RuCl$_3$ powder (Sigma Aldrich, Ru content 45-55\%) sealed in a
quartz ampoule and placed in a three-zone furnace with the end
temperatures 650 and 450$^\circ$C. Those temperatures were chosen
in order to obtain phase-pure \rc (the $\beta$ polytype transforms
irreversibly into the $\alpha$ phase above 395$^\circ$C
[Ref.~\onlinecite{Fletcher67}]) and to ensure that the Cl$_2$ gas
pressure in the ampoule did not exceed atmospheric pressure. The
grown polycrystalline samples contained many flat-plate crystal
pieces, often with a hexagonal shape and up to 1~mm in diameter.
Single crystal x-ray diffraction in the range 80-300~K (under
N$_2$ gas flow) was performed on many of those crystal platelets
using a Mo-source Oxford Diffraction Supernova diffractometer.

Magnetometry measurements were made under static fields using both
a Quantum Design Magnetic Properties Measurement System (MPMS) and
vibrating sample magnetometer (VSM). Pulsed-field magnetization
experiments were performed on a stack of aligned crystal platelets
in both $\mathbf{H} \perp \mathbf{c}^*$ (field in the honeycomb
layers) and $\mathbf{H} \parallel \mathbf{c}^*$ (field
normal to honeycomb layers) geometries. We employed an improved version of
the setup described in Ref.~\onlinecite{Goddard08}, placed within
a $^3$He cryostat with a base temperature of 0.4~K and the 60~T
short-pulse magnet at NHMFL Los Alamos.\cite{Jaime06} The
magnetization values measured in the pulsed-field experiments were
calibrated against VSM data collected on the same sample.

Neutron powder diffraction measurements to obtain information
about the magnetic structure were performed using the
time-of-flight diffractometer WISH at the ISIS Facility in the UK.
Approximately 5~g of powder \rc (extracted from the crystal
growth ampoule described above) was placed in an 
aluminium can and mounted in a standard helium-4 cryostat with a base temperature of 2 K. Additional measurements were performed using a
closed-cycle refrigerator with a base temperature of 6~K.

The electronic structure calculations were performed with the all
electron full potential Wien2k code.\cite{Wien2k} We set the
basis-size controlling parameter $RK_{max}$ equal to 8 and
considered a mesh of $8\times 6\times 8$ ${\bf k}$ points in the
first Brillouin zone (FBZ) for the self-consistency cycle.  The
density of states were calculated with $12\times 12 \times 12$
${\bf k}$ points in the FBZ. All calculations were doubled checked
with the FPLO code.\cite{Koepernik1999}

\section{\label{sec:xrd}Crystal structure}
\subsection{\label{sg}Space group and stacking faults}

The x-ray diffraction pattern was measured for many crystal
platelets extracted from several growth batches. In all samples
studied (over fifty), one could invariably observe sharp
reflections and weak diffuse scattering in rods along the
direction surface normal to the crystal plates, as characteristic
of a layered crystal structure with stacking
faults.\cite{Hendricks} The positions of the sharp Bragg
reflections could be consistently indexed by a monoclinic unit
cell with space group $C2/m$ both at room temperature and the
lowest temperature measured (80~K) with lattice parameters given
in Table~\ref{struc_tab}. Some samples were found to have a single
structural domain, some were found to contain two monoclinic twins
rotated by $\approx$120$^\circ$ about the direction normal to the
plates ($\bf{c}^*$), and other samples contained multiple
structural domains. For the un-twinned crystals the diffraction
patterns had the empirical selection rule for observed Bragg peaks
$h + k = \mathrm{even}$, as characteristic of C-centering in the
$ab$ plane, and the peak intensities were symmetric under a 2-fold
rotation around $\bf{b}^*$ and mirror-plane reflection normal to
$\bf{b}^*$, as expected for a $2/m$ Laue class. The highest symmetry space group consistent with the above information is $C2/m$.

\begin{figure*}[htbp]
\includegraphics[width=1.0\textwidth]{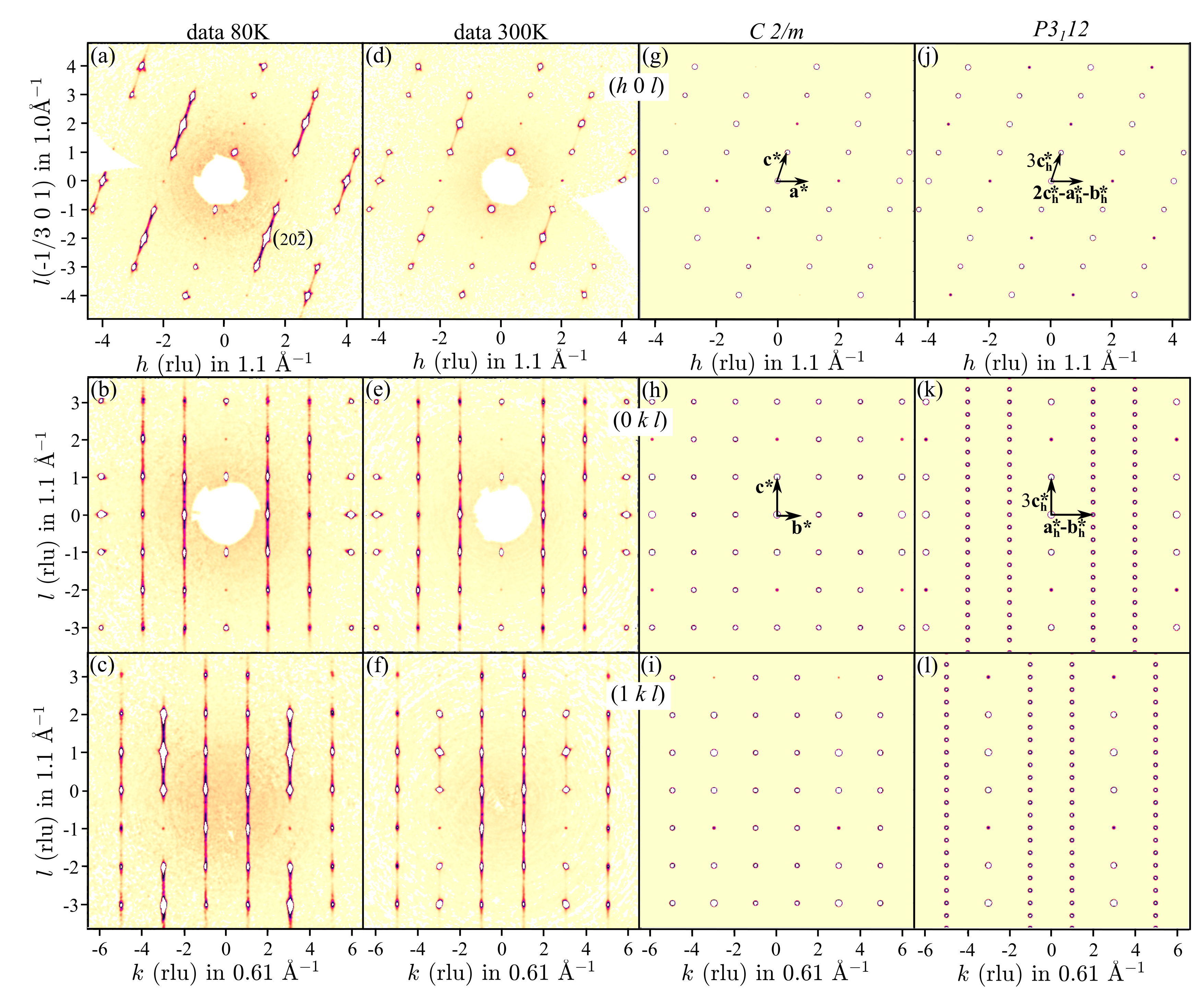}
\caption{(Color online) Observed x-ray diffraction patterns (log
intensity scale) for an un-twinned crystal of \rc at 80~K (A-C),
300~K (D-F), shown for three different planes, compared with
calculations (G-I) for the monoclinic $C2/m$ structural model
(Fig.~\ref{fig:structure}) and the trigonal $P\,3_112$ model
(J-L). All wavevectors are labelled in r.l.u. units of the
monoclinic cell and ${\bf a}_h^*$, ${\bf b}_h^*$ and ${\bf c}_h^*$
denote reciprocal lattice vectors of the hexagonal primitive cell
of the trigonal structure (for the relation between the hexagonal
and monoclinic axes see Sec.~\ref{sc}). Note the sharp peaks in
the data are in good agreement with the monoclinic model (compare
D-F with G-I), whereas the ``supercell" peaks expected in the case
of the trigonal model (K-L) at fractional positions $l=n+1/3,
n+2/3$ ($n$ integer) are clearly absent from the data, instead
only diffuse scattering is found in those places. }
\label{fig:comparison}
\end{figure*}

Representative data at 300~K from an un-twinned crystal (of
$\approx80\mu$m diameter) is shown in
Fig.~\ref{fig:comparison}D-F, for various diffraction planes. Note
that all sharp Bragg peaks are in good agreement with calculations
(panels G-I) for a $C2/m$ structure. In addition to sharp Bragg
peaks, rods of diffuse scattering are also clearly visible along
$l$ (see panels E-F), with the general selection rule $k=3n+1$ or
$3n+2$ ($n$ integer) and $h+k=\mathrm{even}$ (due to C-centering).
Diffuse scattering with the same selection rule was also observed
in \nio and attributed to faults in the stacking sequence of
honeycomb Na$_{1/2}$IrO$_3$ layers.\cite{choi12} By analogy, we
attribute the above diffuse scattering observed in \rc as
originating from occasional shifts in the $ab$ plane by $\pm{\bf
b}/3$ between stacked RuCl$_3$ honeycomb layers. The intensities
of the sharp Bragg peaks located at integer $l$ positions on those
diffuse scattering rods are expected to have a reduced intensity
compared to a fully-ordered structure due to some transfer of
intensity into the diffuse rod.\cite{Hendricks} For the
quantitative structural refinement we will show that it is helpful
to distinguish between different families of Bragg peaks, and for
this purpose we label the above family of Bragg peaks whose
intensities are affected by diffuse scattering from sliding
stacking faults as in \nio, as `SFa' (peaks affected by Stacking
Faults of type `a', to distinguish them from another family of
type `b', to be discussed below).

Upon cooling to low temperatures (80~K) no new diffraction peaks
appear, but a second family of diffuse scattering rods becomes
apparent. This is most clearly seen by comparing
Figs.~\ref{fig:comparison}A and D, note the diffuse scattering rod
near (2,0,-2) (panel A, labelled peak position), which is
prominent at low temperature, but only just visible at 300 K
(panel D). Note also in Fig.~\ref{fig:comparison}C the strong
diffuse scattering along $(1,\pm3,l)$ positions, almost absent at
300 K (panel F). This type of diffuse scattering was not detected
at 300 K in \nio [Ref.~\onlinecite{choi12}] and has a different
selection rule ($k=3n$ and $h=3m+1$ or $3m+2$ with $n$, $m$
integers and $h+k=$ even) compared to the diffuse scattering of
type `a' discussed previously. At 80~K both families of diffuse
scattering rods have comparable intensities (see
Fig.~\ref{fig:comparison}C). As before, the Bragg peaks located at
integer $l$ positions on this second family of diffuse scattering
rods are expected to be reduced in intensity compared to a
fully-ordered structure. We label this family of Bragg peaks as
`SFb' (peaks affected by Stacking Faults of type `b').

Finally, a third family of Bragg peaks exist that are sharp at all
temperatures measured, such as $(00n)$ ($n$ integer) in
Fig.~\ref{fig:comparison}A, so appear not to be affected by the
presence of stacking faults. These have the general reflection
condition $h=3m$ and $k=3n$ ($n,m$ integers and $h+k=$ even), and
we label them `NSF' (peaks Not affected by Stacking Faults).

\begin{table}[htb]
\caption{\label{struc_tab}\rc crystal structure parameters at
80~K.}
\begin{ruledtabular}
\begin{tabular}{c c c c c}
\multicolumn{4}{l}{\textbf{Cell parameters}} \\
\multicolumn{4}{l}{Space group: $C2/m$} \\
\multicolumn{4}{l}{$Z = 4$} \\
\multicolumn{2}{c}{$a,b,c$ ({\AA})} & 5.9762(7) & 10.342(1) & 6.013(1) \\
\multicolumn{2}{c}{$\alpha,\beta,\gamma$ ($^\circ$)} & 90 & 108.87(2) & 90 \\
\multicolumn{2}{c}{Volume ({\AA}$^3$)} & 371(2)& & \\
\\
\multicolumn{4}{l}{\textbf{Atomic fractional coordinates from DFT}} \\
Atom & Site & $x$ & $y$ & $z$ \\
\hline
Ru & 4g & 0 & 0.33441 & 0 \\
Cl1 & 8j & 0.75138 & 0.17350 & 0.76619 \\
Cl2 & 4i & 0.73023 & 0 & 0.23895 \\
\\
\multicolumn{4}{l}{\textbf{Selected bond lengths and angles from DFT}} \\
\multicolumn{2}{c}{Ru$_1$-Ru$_2$} &  \multicolumn{2}{c}{3.42513 {\AA}}\\
\multicolumn{2}{c}{Ru$_2$-Ru$_3$} &  \multicolumn{2}{c}{3.46080 {\AA}}\\
\multicolumn{2}{c}{Ru$_1$-Cl2-Ru$_2$} &  \multicolumn{2}{c}{92.5954$^\circ$} \\
\multicolumn{2}{c}{Ru$_2$-Cl1-Ru$_3$} &  \multicolumn{2}{c}{93.9310$^\circ$} \\
\\
\multicolumn{5}{l}{\textbf{Fitted isotropic atomic displacement parameters}} \\
\multicolumn{2}{c}{Atom} & \multicolumn{2}{c}{$U_{iso}$({\AA}$^2$)} \\
\hline
\multicolumn{2}{c}{Ru} & \multicolumn{2}{c}{0.005(1)} \\
\multicolumn{2}{c}{Cl1} &  \multicolumn{2}{c}{0.006(2)} \\
\multicolumn{2}{c}{Cl2} &  \multicolumn{2}{c}{0.006(2)} \\
\\
\multicolumn{4}{l}{\textbf{Data collection}} \\
& & SFa & SFb & NSF\\
\hline
\multicolumn{2}{c}{$\#$ measured refl.} & 991 & 325 & 135 \\
\multicolumn{2}{c}{$\#$ independent refl.} & 189 & 68 & 32 \\
\multicolumn{2}{c}{$R_\mathrm{int}(C2/m)$} & 8.0\% & 3.3\% & 2.9\% \\
\\
\multicolumn{4}{l}{\textbf{Fit to NSF peaks}} \\
\multicolumn{4}{l}{(Criterion for observed reflections: $I > 3.0\sigma(I)$)} \\
\multicolumn{4}{l}{$\#$ observed reflections: 32} \\
\multicolumn{4}{l}{$\#$ fitted parameters: 3} \\
\end{tabular}
\end{ruledtabular}
\end{table}

\subsection{\label{sr}Structural refinement at 80 K}
\begin{figure}[htbp]
\includegraphics[width=0.48\textwidth]{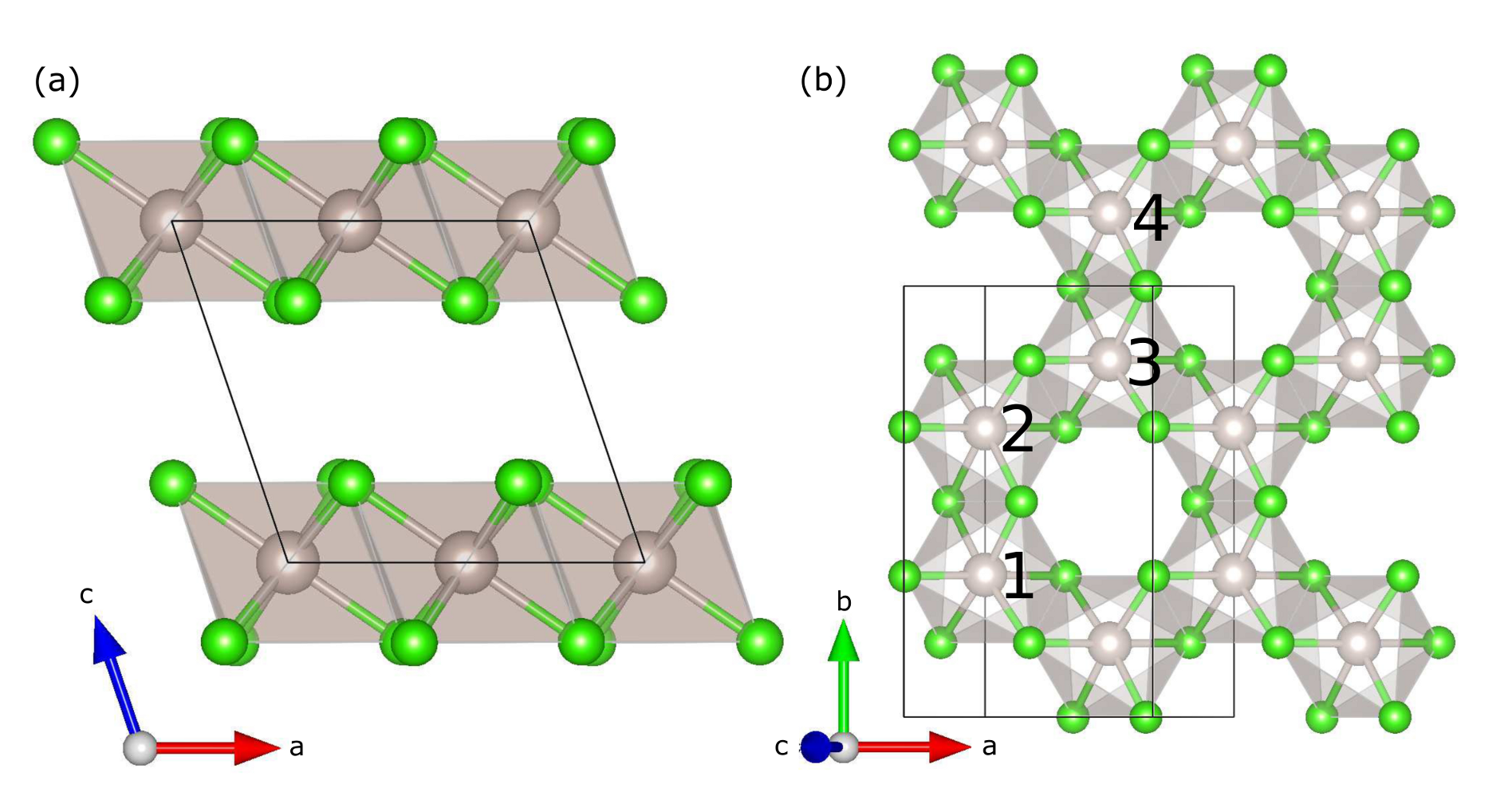}
\caption{\label{fig:structure}(Color online) Monoclinic crystal
structure of $\alpha$-RuCl$_3$, showing the unit cell as a black
outline, Ru as grey balls and Cl as green.(a) Projection onto the
$ac$ plane. (b) Basal layer projected onto the $ab$ plane.}
\end{figure}

To obtain a reference, fully-ordered 3D structure with no stacking
faults we must refine a structural model against only those
diffraction peaks that are unaffected by the presence of stacking
faults. These are the family labelled NSF, as defined above. In
the following, we focus primarily on the data collected at 80~K.
Out of a total 1451 Bragg peaks measured, 135 are NSF peaks, of
those just 32 are symmetry inequivalent after data reduction in
space group $C2/m$. Despite the small number of reflections a full
refinement using FULLPROF\cite{rodriguezcarvaja93} of a structural
model, with starting atomic positions for Ru and Cl taken to be
those of Ir and O in the structure of \nio, converged well. Hence,
the data was found to be fully consistent with the same structural
motif as that found in \nio with honeycomb layers of edge-sharing
RuCl$_6$ octahedra stacked vertically with an in-plane offset (see
Fig.~\ref{fig:structure}), with Ru in place of Ir, Cl in place of
O, and removing Na altogether. However, detailed tests showed that
the refinement was in fact not sufficiently sensitive to the
$y$-position of the Ru ion, or the precise distortions of the
Cl$_6$ octahedra, so the internal atomic fractional coordinates
could not be uniquely determined from the x-ray data alone. The
atomic positions are key to understanding the underlying physics
as the exchange interactions (and their anisotropy) are expected
to be strongly dependent on the geometry of the Ru-Cl-Ru bonds. So
to construct a robust structural model we use {\em ab-initio}
density functional theory (DFT) calculations to predict the atomic
positions that give the lowest energy ground state using as input
the experimentally determined space group and lattice parameters,
and then check consistency of this constrained structural model
with the intensities in the x-ray diffraction data. For the DFT
structural relaxation calculations we employed the projector
augmented wave method as implemented in the VASP
package\cite{VASP} with the generalized gradient approximation
(GGA)\cite{Perdew1996}, as well as the full potential local
orbital (FPLO) method.\cite{Koepernik1999}

\begin{figure}[htbp]
\includegraphics[width=0.48\textwidth]{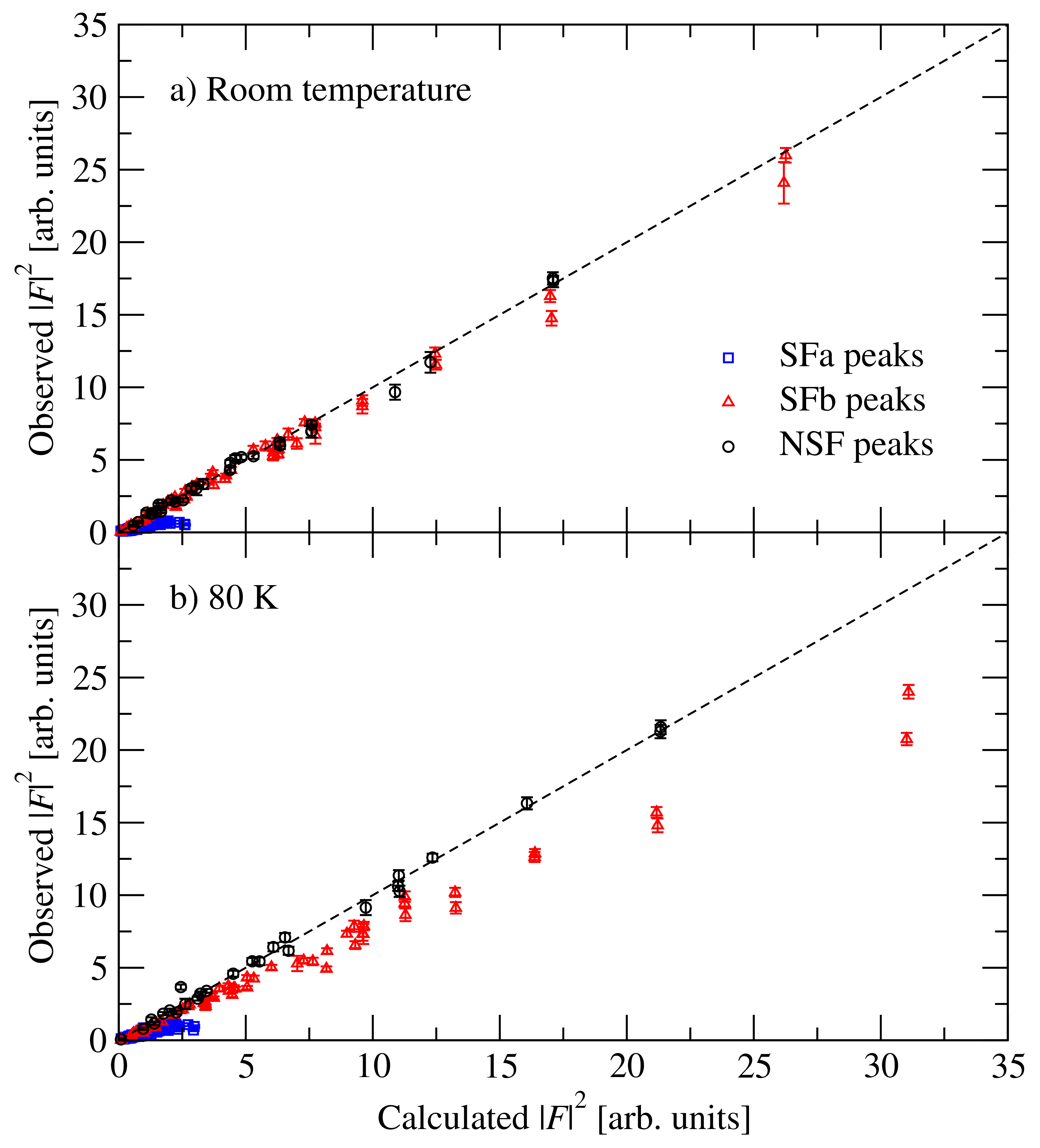}
\caption{\label{fcomp_fig}(Color online) Observed structure factor
squared values of all three families of diffraction peaks compared
to those calculated by fitting the $C2/m$ monoclinic structural
model with fixed theoretical atomic fractional coordinates to data
measured at a) room temperature, and b) 80~K.}
\end{figure}

The atomic fractional coordinates predicted by DFT within the
above empirical constraints are given in Table~\ref{struc_tab}.
The refinement of the structural model against the 80~K NSF peak
intensities was repeated with atomic fractional coordinates fixed
to those DFT values, with only isotropic displacement parameters
and a global scale factor left free to vary. A reliability factor
of $R_{F^2} = 4.2 \%$ was obtained, which compared to a value of
$R_{F^2} = 3.7 \%$ achieved for the completely free refinement
(when atomic coordinates were also allowed to vary), demonstrates
that the theoretically predicted atomic coordinates are fully
consistent with the x-ray diffraction data. Fig.~\ref{fcomp_fig}
shows the observed structure factors squared, $|F|^2$, for all
families of diffraction peaks compared to those calculated from
the fit against only the NSF peaks. The excellent agreement with
the NSF peak intensities at 80~K is clear (Fig.~\ref{fcomp_fig}b,
black symbols). Furthermore, one can see that intensities of both
SFa (blue) and SFb (red) peaks are systematically overestimated,
consistent with the expectation that some of their nominal
intensity has been transferred into the diffuse scattering in
their vicinity. Fig.~\ref{fcomp_fig}a shows the same fit, but
performed against the room temperature data set (with empirically-
determined lattice parameters $a=5.9856(4)$\AA, $b=10.3557(5)$\AA,
$c=6.0491(4)$\AA, $\beta=108.828(7)^\circ$ and assuming  atomic
fractional coordinates fixed to the DFT predicted values listed in
Table~\ref{struc_tab}). Even at this temperature, the structural
model agrees well with the x-ray data ($R_{F^2} = 5.5 \%$ for NSF
peaks), and the intensities of the SFb peaks (red symbols) appear
to be also almost quantitatively reproduced by the model, as at
this temperature the diffuse scattering near SFb peaks is almost
absent, so the intensity of SFb peaks is expected to be only very
weakly reduced compared to a perfectly-ordered structure.

The obtained crystal structure allows us to naturally understand
the strong periodic modulations in the intensity of x-ray
diffraction peaks, in particular the rather conspicuous period-4
repeat in the intensity of peaks along $l$ in the $(h0l)$ plane
(see Fig.~\ref{fig:comparison}D) with almost extinct peaks at
$(0,0,\pm2)$ and $(\pm2,0,0)$ positions. The near-absence of
intensity at those positions is due to an almost total
cancellation of the scattering from the Ru ion with that from the
three Cl ions with atomic scattering amplitudes
$f_\mathrm{Ru}:f_\mathrm{Cl}$ in ratio almost $3:1$. In detail,
the Ru ion is located at $z=0$, whereas the Cl ions are at $z
\simeq 1/4$ and 3/4 (see Table~\ref{struc_tab}), so the structure
factors for $(00l)$ reflections follow (to a good approximation) a
period-4 sequence of values $f_\mathrm{Ru}+3f_\mathrm{Cl}$,
$f_\mathrm{Ru}$, $f_\mathrm{Ru}-3f_\mathrm{Cl}$, $f_\mathrm{Ru}$,
$f_\mathrm{Ru}+3f_\mathrm{Cl}\ldots$. In the limit of small
wavevectors $\bf{Q}$, the atomic scattering factors are given by
the number of electrons, so $f_\mathrm{Ru}:f_\mathrm{Cl} = 44:17
\approx 3:1$, such that to good approximation the structure
factors are multiples of $4, 1, 0, 1, 4 \ldots$ for $l=0,1,2,3,4
\ldots$. Similarly, one can show that a period-4 modulation in
intensity along $l$ occurs in general for $(h0l)$ peaks, with
zeros at $h+l=4n+2$ ($n$ integer and $h$ even) explaining all the
near-extinctions and apparent intensity modulations seen in
Fig.~\ref{fig:comparison}D. We note that such near extinctions do
not occur in the diffraction pattern of the iso-structural \nio,
as the scattering factors of Ir and O are much more anisotropic
(ratio almost 10:1) and Na is also contributing to the diffraction
peak intensities.

To summarize, the x-ray diffraction patterns uniquely identify the
monoclinic $C2/m$ space group both at room temperature and the
lowest temperature measured (80~K), and quantitative structural
refinement using fixed atomic fractional coordinates predicted by
DFT, performed only against the sharp diffraction peaks whose
intensity is not affected by the presence of stacking faults,
gives a very good description of the data. The corresponding
crystal structure is shown in Fig.~\ref{fig:structure} and
consists of monoclinically-stacked RuCl$_3$ honeycomb layers as in
AlCl$_3$ [Ref.~\onlinecite{Ketelaar}] and \nio. The real materials
are understood to have occasional stacking faults with respect to
this reference monoclinic structure.

\subsection{\label{sc}Other Structural Models}
The current structural model assumed for \rc (trigonal space group
$P\,3_112$ [Ref.~\onlinecite{stroganov57}], conventionally
described in a hexagonal unit cell) differs from the monoclinic
$C2/m$ structure primarily in the stacking sequence of the
honeycomb layers, with a three-layer stacking periodicity as
opposed to single layer in $C2/m$. We note that the dimensions of
the unit cell are, in general, an insufficiently robust criterion
to reliably distinguish between those two structural models as the
monoclinic unit cell metric is in fact very close to hexagonal,
i.e. $b\simeq\sqrt{3}a$ to within better than 0.2\%, and
$3c\times\cos\beta\simeq-a$ to within 2\%. When the latter
equation is satisfied exactly one has eclipsed (straight-on-top)
stacking at the $3^\mathrm{rd}$ honeycomb layer, so an alternative
hexagonal cell with a 3-layer periodicity along the direction
normal to the layers could in principle provide an approximate
metric to index the positions of Bragg diffraction peaks. In this
case the transformation between the hexagonal (subscript $h$) and
symmetrized monoclinic unit cell vectors (subscript $m$) is given
by ${\mathbf a}_m=-{\mathbf a}_h-{\mathbf b}_h$, ${\mathbf
b}_m={\mathbf a}_h-{\mathbf b}_h$, ${\mathbf c}_m=({\mathbf
a}_h+{\mathbf b}_h+{\mathbf c}_h)/3$, where $a_m = a_h$, $b_m =
\sqrt{3}a_h$, $\beta =\pi/2+\mathrm{atan}(a_h/c_h)$, and
$c_m=c_h/(3\sin\beta)$.

However, the internal atomic arrangement in the monoclinic and
trigonal structures is different due to the distinct symmetries of
the corresponding space groups, and these differences would be
directly observed in the measured single crystal diffraction
patterns. In particular, the two structures have a distinct
stacking sequence of the honeycomb layers: for two adjacent layers
both the symmetrized monoclinic and trigonal structures would
appear identical, but for every subsequent layer in the trigonal
structure the direction of the in-plane offset (defined by the
monoclinic angle, $\beta$) would rotate by 120$^\circ$ around the
direction normal to the layers. The resulting 3-layer stacking
periodicity in the trigonal structure would lead to the appearance
of extra \emph{supercell} peaks along the $\mathbf{c}^*$-axis,
which, in the monoclinic basis, would occur at non-integer
positions $l=n+1/3$ and $n+2/3$ ($k=3m+1$ or $3m+2$, and $h+k=$
even with $h$,$m$,$n$ integers) in addition to, and with the same
intensity as, the nominal peaks at integer $l=n$ positions (see
Fig.~\ref{fig:comparison}K-L). The absence of supercell peaks in
our diffraction data [compare Fig.~\ref{fig:comparison}E-F with
K-L) conclusively rules out the proposed $P\,3_112$ model. For
completeness, we note that an alternative rhombohedral stacking
sequence of the honeycomb layers with space group $R\bar{3}$
proposed\cite{kubota15} for \rc by analogy with the
low-temperature phase of CrCl$_3$ [Ref.~\onlinecite{Morosin}],
also has a 3-layer stacking periodicity so would also predict
supercell peaks at non-integer $l=n+1/3$ and $n+2/3$ positions,
not observed in the data, so this rhombohedral structure can
similarly be ruled out for the crystals studied here.

We note that if a sample contained three monoclinic twins of equal
weight and rotated by 120$^{\circ}$ around $\bm{c}^*$, then there
would be no striking qualitative difference between the
diffraction pattern from monoclinic and trigonal/rhombohedral
structural models. Furthermore, under the symmetry constraints of
those candidate structures there would be only slight variations
in intensity due to differences in the displacements of the Cl or
Ru ions from their idealized positions, which are expected to be
small and likely below the experimental sensitivity. As such,
measuring un-twinned crystals has proved to be crucial in the
present study to qualitatively, and quantitatively, determine the
correct monoclinic reference structure for the samples reported
here.

\section{\label{sec:magnetism}Magnetic Properties}

\subsection{\label{magnetization}Susceptibility and magnetization}

\begin{figure}[htbp]
\includegraphics[width=0.48\textwidth]{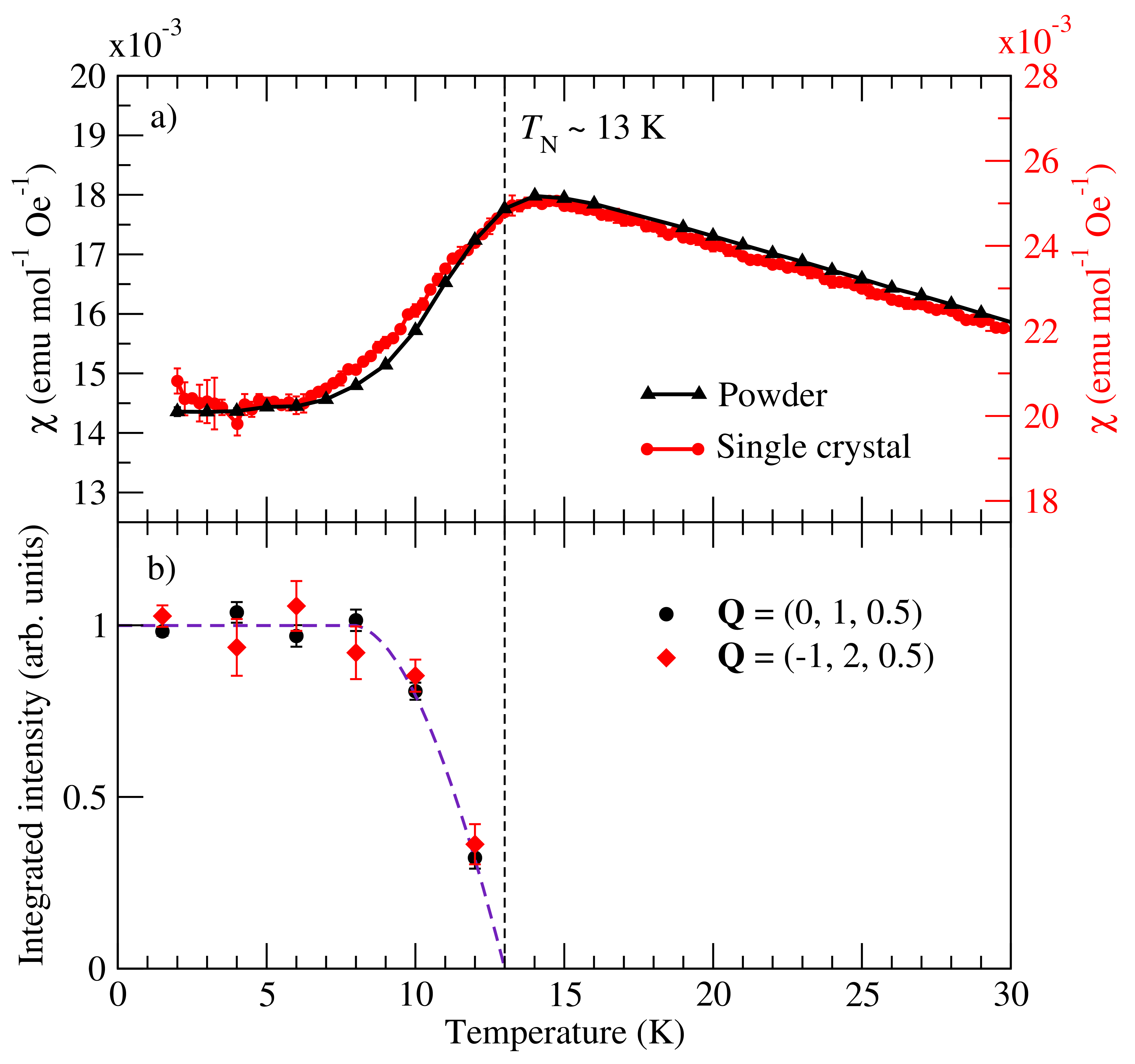}
\caption{\label{fig:tempdep}(Color online) a) Magnetic
susceptibility as a function of temperature for a stack of single
crystals (red circles, ${\mathbf H}\perp \mathbf{c}^*$)
representative of those used in the x-ray diffraction experiments
described in Sec.~\ref{sec:xrd}, and the powder sample used in the
neutron diffraction experiments discussed in Sec.~\ref{npd} (black
triangles), in a magnetic field $H=1000$~Oe. b) Temperature
dependence of the integrated intensity of the two magnetic
reflections observed in neutron powder diffraction pattern in
Fig.~\ref{fig:npd}, normalized to an average of unity at low
temperatures. The dashed line is a guide to the eye.}
\end{figure}

The magnetic susceptibility of a stack of single crystals
representative of those used in our structural study
(Sec.~\ref{sec:xrd}), and a 12.8~mg powder, was measured on
heating (after zero-field cooling) from 2~K up to 300~K. Only a
single anomaly was observed for both samples near 13~K [see
Fig.~\ref{fig:tempdep}a], which is indicative of long-range
antiferromagnetic ordering of the ruthenium magnetic moments. Our
powder data is fully consistent (in absolute units) with data
previously reported on powder \rc samples.\cite{kobayashi92}
Previous single crystal studies have reported two magnetic
transitions near 8 and 14~K
[Refs.~\onlinecite{majumder15,sears15,kubota15}], which have been
attributed to either a mixture of two coherent stacking orders,
with each order associated with a single transition,
respectively,\cite{Banerjee15} or alternatively to a single phase
that supports an unexpected magnetic ground state.\cite{sears15}
Here, to the contrary, we find that the low-field magnetic
susceptibility of single crystals is consistent with that of the
powder, both displaying a single transition to magnetic order at
low temperatures.

\begin{figure*}[htbp]
\includegraphics[width=1.0\textwidth]{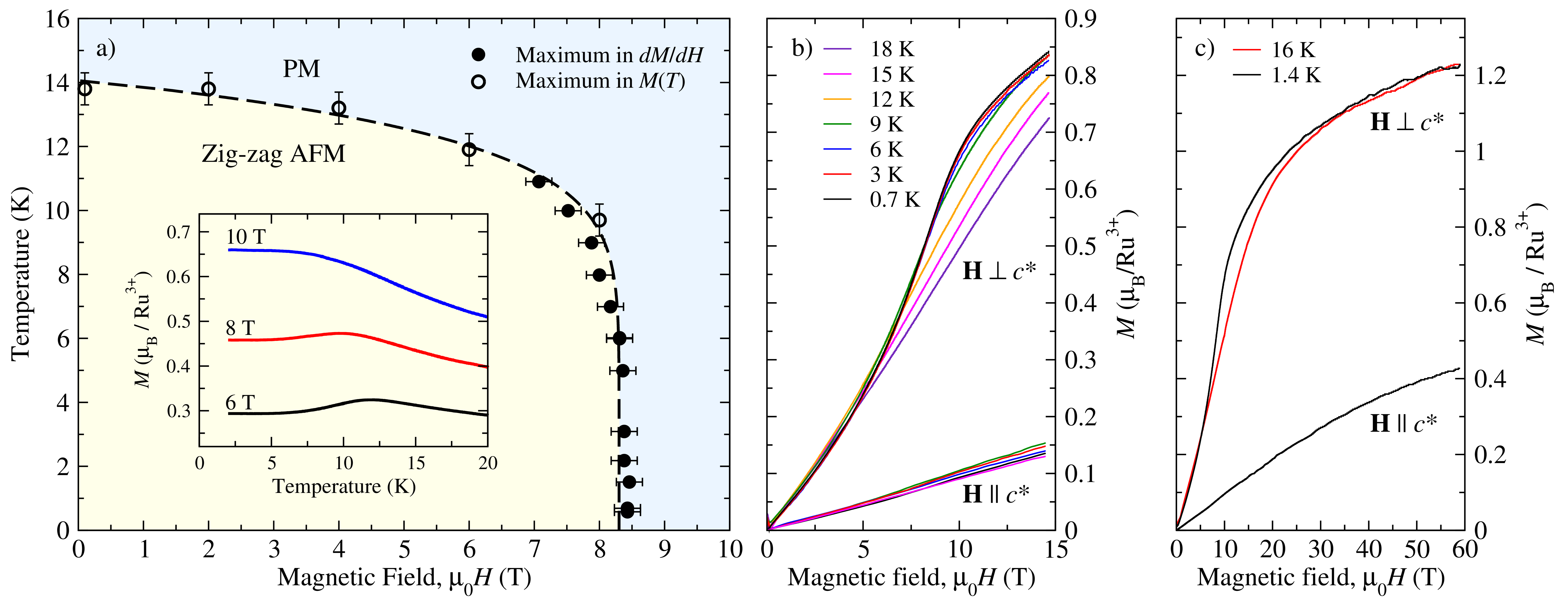}
\caption{\label{fig:pulsed_field}(Color online) a) Magnetic phase
diagram for single crystal \rc in magnetic field $\mathbf{H}\perp
\mathbf{c}^*$. Solid points mark the maxima in the differential
susceptibility ${\rm d}M/{\rm d}H$ derived from data shown in
panel b) (upper traces). Open symbols mark the maximum in $M(T)$
VSM temperature sweeps, as shown in the pane inset for constant
magnetic field values close to the phase boundary. The dashed line
is a guide to the eye phase boundary between the zigzag
antiferromagnetic phase (yellow shading) and paramagnetic (PM,
blue shading). b)~$M(H,T)$ data recorded in the rising part of
15~T field pulses at a series of constant temperatures. At lower
temperatures, the steep rise in $M(H)$ is strongly suggestive of a
field-induced phase transition near 8~T. c)~$M(H,T)$ data recorded
in the rising part of 60~T field pulses in both the
antiferromagnetic and paramagnetic phases.}
\end{figure*}

Pulsed-magnetic-field $M(H,T)$ data are shown for field sweeps up
to 15~T at various constant temperatures $T$ in
Fig.~\ref{fig:pulsed_field}b. The data shown was recorded during
the rising part of the field pulses; $M(H)$ curves from the rising
and falling portions of the field pulse were indistinguishable
within the limit of experimental sensitivity ({\em i.e.} there was
little or no hysteresis). For $\mathbf{H}\perp \mathbf{c}^*$ the
low-temperature $M(H)$ curves show a pronounced steepening at
about 8~T, characteristic of a field-induced phase transition,
which gradually shifts down in field and fades as the temperature
increases. This trend is more clearly seen in the full phase
diagram shown in Fig.~\ref{fig:pulsed_field}a), which displays
maximum values (solid symbols) of the differential susceptibility
(${\rm d}M/{\rm d}H$) as a function of $H$ and $T$. The inset to
Fig.~\ref{fig:pulsed_field}a) shows complementary $M(H,T)$ data
recorded in the VSM as temperature sweeps in fixed field. The same
transition is seen as a peak in $M(T)$ that disappears at fields
above 8~T. This trend is also drawn in the main pane of
Fig.~\ref{fig:pulsed_field}a), which completes a continuous phase
boundary (dashed line) consistent with a single enclosed
antiferromagnetic phase for \rc at low temperatures and modest
magnetic fields applied in the honeycomb layers.

The pulsed-field data shown in Fig.~\ref{fig:pulsed_field}b) for
$\mathbf{H}\parallel \mathbf{c}^*$ exhibit $M(H,T)$ values that
are a factor $5-6$ times smaller than those recorded on the same
sample under comparable conditions for $\mathbf{H}\perp
\mathbf{c}^*$. This is likely to be due to Ru $g$-factor
anisotropy.\cite{kubota15} Note that there is no sign of the phase
transition observed in the other field orientation, leading us to
conclude that it is a feature observed only when the field lies in
the honeycomb plane.

Having measured the magnetization along the two non-equivalent
directions on the {\em same} sample enables us to reliably put
both data sets in absolute units by calibration against the
susceptibility data measured on a powder sample
[Fig.~\ref{fig:tempdep}a) black symbols] under the same conditions
of applied field and temperature, thus avoiding the inherent
uncertainties associated with measuring the precise mass of very
small (of order $\sim$0.1~mg) crystals. The powder susceptibility
is expected to reflect the spherically-averaged value, obtained as
$\chi_{\rm
powder}=(2/3)\chi_{\parallel}+(1/3)\chi_{\perp}=\chi_{\parallel}(2+r)/3$,
where $r=\chi_{\perp}/\chi_{\parallel}$ is the susceptibility
anisotropy. The single crystal data sets in
Figs.~\ref{fig:tempdep}a) (red symbols) and
Fig.~\ref{fig:pulsed_field}b-c) were then scaled to satisfy the
above relations with the powder susceptibility data at
$\mu_0H=0.1$ T and 15~K, where the susceptibility anisotropy under
those conditions was obtained as $r=0.157$ from the pulsed field
data.

Fig.~\ref{fig:pulsed_field}c) shows $M(H,T)$ data recorded in 60~T
pulsed-field shots; as is the case with the lower-field data,
there is little or no hysteresis between up- and down-sweeps of
the field, and so, for clarity, only data recorded on the rising
part of the field pulse are shown. The $M(H)$ anisotropy persists
to high fields, though the data for $\mathbf{H}\perp \mathbf{c}^*$
show signs of the approach to saturation. There are no further
phase transitions visible up to 60~T in either field direction.

The shape of the magnetization curve at high field as observed by
the upper traces in Fig.~\ref{fig:pulsed_field}c) with a gradually
decreasing differential susceptibility upon increasing field
suggests an asymptotic approach to magnetization saturation. Such
a behaviour of the magnetization near saturation is commonly
seen\cite{Kenzelmann,Trebst} when the spin Hamiltonian does not
have rotational symmetry around the applied field direction. In
this case the total spin along the field direction
$S_\mathrm{T}^\xi=\sum_i S_i^\xi$ is not a good quantum number
(the operator does not commute with the spin Hamiltonian
$[S_\mathrm{T}^{\xi},{\cal H}] \neq 0$, where $\xi$ denotes the
direction of the applied field $\mathbf{H}$ and $i$ runs through
all the magnetic sites) and as a consequence even in the limit of
very high fields quantum fluctuations are still present and reduce
the magnetization from its fully-available value, with saturation
strictly reached only in the asymptotic limit of infinite field.
This is qualitatively different from the case when the spin
Hamiltonian does have rotational symmetry around the field
direction, for example the case of purely Heisenberg interactions,
${\cal H}=\sum_{ij}J_{ij}\mathbf{S}_i\cdot\mathbf{S}_j$. In this
case the total spin along the field direction is a good quantum
number, magnetization saturation is an exact plateau phase where
quantum fluctuations are entirely absent, and the approach to
magnetization saturation from below is via a sharp phase
transition at a critical field $H_C$, with the susceptibility in
general increasing upon increasing field up to $H_C$, then being
strictly zero above it. The observed shape of the magnetization
curve at high field (upper traces in Fig.~\ref{fig:pulsed_field}c)
is consistent with the former scenario with an asymptotic approach
to saturation and could be taken as evidence for the presence of
strongly anisotropic, non-Heisenberg exchanges in \rc, of
Kitaev\cite{Trebst} or another strongly-anisotropic form.

\subsection{\label{npd}Magnetic Neutron Powder Diffraction}

\begin{figure}[htbp]
\includegraphics[width=0.48\textwidth]{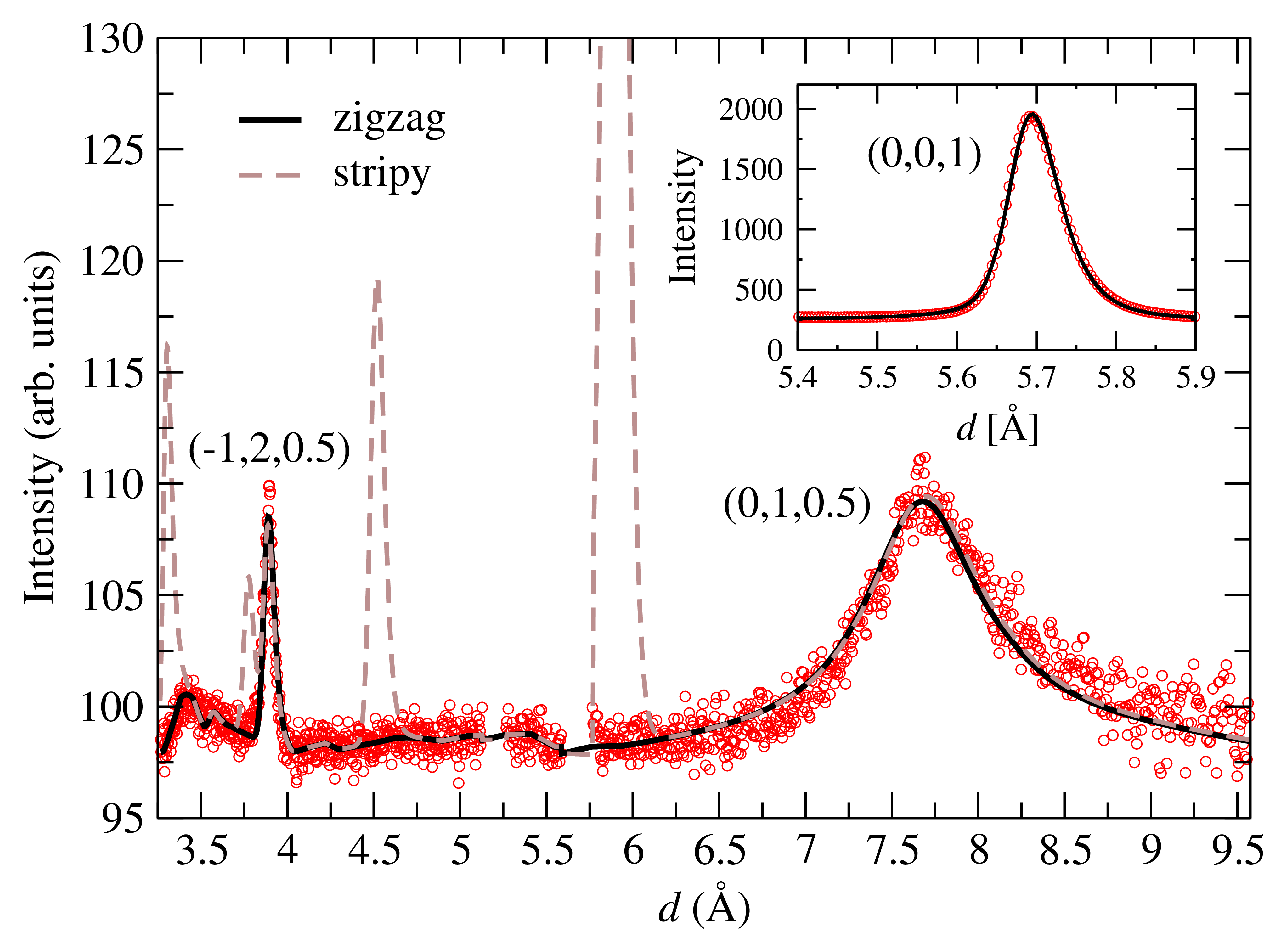}
\caption{\label{fig:npd}(Color online) Neutron powder diffraction
data measured at 6~K, with the 20~K paramagnetic data subtracted.
The diffraction pattern for both zigzag (black solid line) and
stripy (brown dashed line) models are calculated and plotted for a
moment oriented along $\mathbf{c}^*$, a similar level of agreement
for the zigzag structure could be obtained for a general moment
direction in the $ac$ plane. Inset: fit to the $(001)$ nuclear
Bragg reflection, unaffected by stacking faults, used for
calibrating the magnetic diffraction intensities.}
\end{figure}

Neutron powder diffraction data were collected deep in the ordered
phase (6~K) and in the paramagnetic region (20~K) with high
counting statistics to allow a quantitative refinement. Additional
data to monitor the temperature dependence and extract an order
parameter was collected with lower statistics at 2~K intervals in
the range 2-14~K. Fig.~\ref{fig:npd} shows the purely magnetic
contribution to the neutron diffraction pattern at 6~K obtained
after subtracting off the 20~K paramagnetic pattern. Two clear
magnetic diffraction reflections are observed at $d$-spacings $d =
3.88$ and 7.67 {\AA}. The integrated intensity of the two
reflections is plotted as a function of temperature in
Fig.~\ref{fig:tempdep}b. Both peaks show the same temperature
dependence, and clearly demonstrate the onset of long-range
magnetic order below $T_\mathrm{N}\approx13$~K. Furthermore, both
magnetic susceptibility and neutron diffraction data are
consistent with a single magnetic ordered phase down to the lowest
temperature measured (2~K).

\begin{figure}[htbp]
\includegraphics[width=0.48\textwidth]{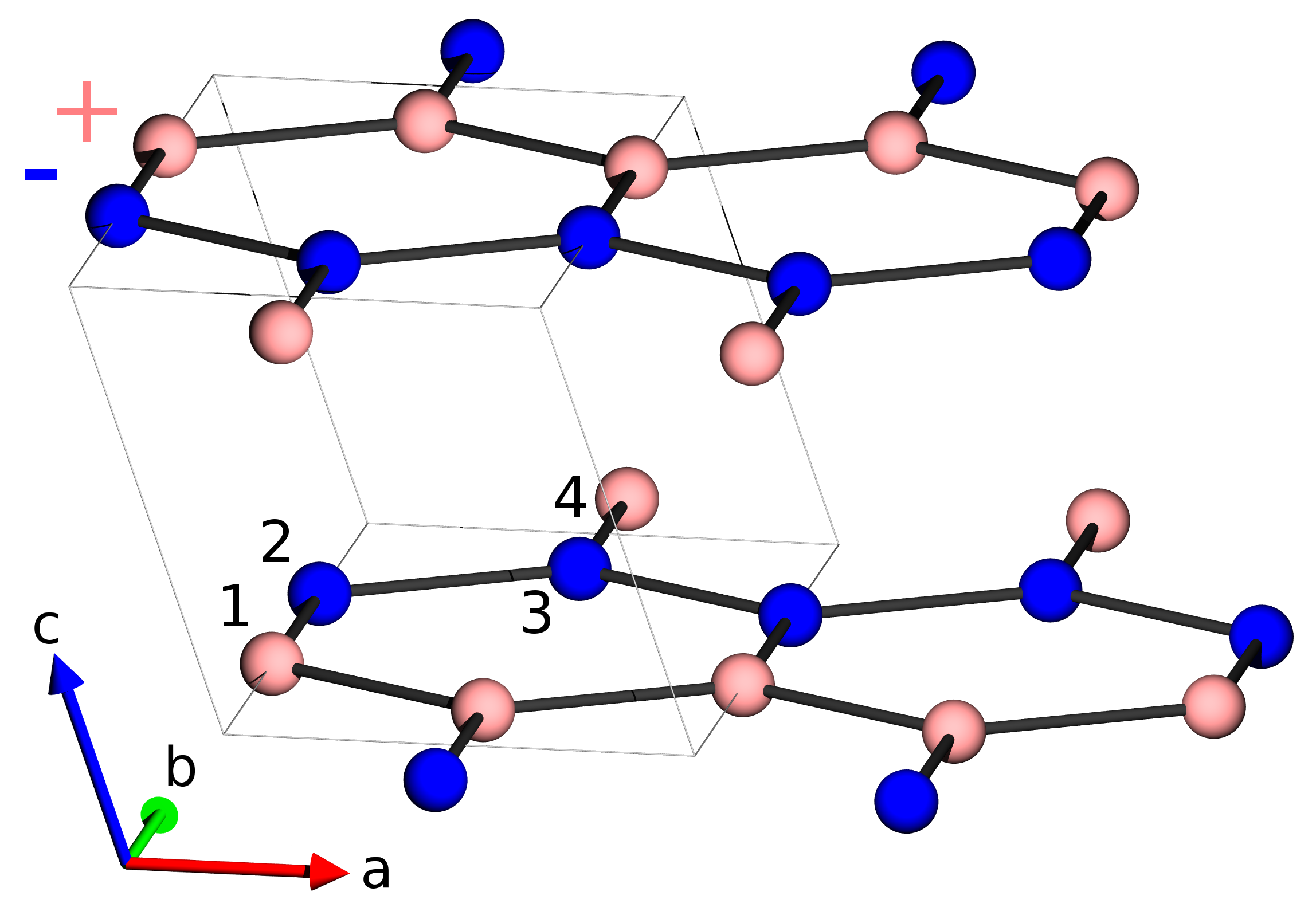}
\caption{\label{fig:magstructure}(Color online) The zigzag
magnetic structure of \rc. The magnetic moments of ruthenium atoms
colored red and blue are aligned antiparallel and oriented within
the $ac$-plane. Ru-Ru connections are drawn in thick black lines
to illustrate the honeycomb layers, and the $C2/m$ monoclinic unit
cell is drawn in thin gray lines.}
\end{figure}

Both magnetic reflections could be indexed with the propagation
vector ${\mathbf k}=(0,1,0.5)$ with reference to the $C2/m$
structural unit cell. This finding alone provides key information
on the ground state magnetic structure of our \rc samples. The
value $k_z = 0.5$ determines that the magnetic moments in
neighbouring honeycomb layers are aligned antiferromagnetically.
Within a honeycomb layer there are four symmetry equivalent
ruthenium ions per unit cell, labelled 1-4 in
Fig.~\ref{fig:structure}b) and Fig.~\ref{fig:magstructure}. The
four sites can be considered as two pairs of sites, (1 and 2) and
(3 and 4), intra-related by mirror symmetry operations at
$(x,\tfrac{1}{2},z)$ and $(x,0,z)$, respectively, and
inter-related by the $C$-centering translation vector ${\mathbf
t}=(\tfrac{1}{2},\tfrac{1}{2},0)$. The relative orientation of the
magnetic moment pairs, (1 and 2) and (3 and 4), is uniquely
determined by the phase $2\pi\mathbf{k}\cdot\mathbf{t}$,
\textit{i.e.} for $k_x=0$ and $k_y=1$ the two pairs are aligned
antiferromagnetically. Furthermore, for this $\mathbf{k}$-vector
the relative orientation of moments within a given pair is
strictly parallel or antiparallel by symmetry, however these two
scenarios are not differentiated by the propagation vector alone
and must be tested against the diffraction data. For parallel
alignment within each pair the resulting magnetic structure is a
`stripy' antiferromagnet with spins forming ferromagnetic stripes
(ladders) along $a$ alternating in orientation along $b$. In the
case of antiparallel alignment within each pair the magnetic
structure consists in `zigzag' ferromagnetic chains along $a$
arranged in an antiferromagnetic pattern along $b$, as illustrated
in Fig.~\ref{fig:magstructure}. Symmetry analysis performed using
BasIreps, part of the FULLPROF package,\cite{rodriguezcarvaja93}
for the propagation vector ${\bf k}$ gives magnetic basis vectors
containing moments aligned along the $b$-axis (the unique 2-fold
axis of the crystal structure) or in the $ac$ plane. If the
transition from paramagnetic to magnetic order is continuous then
the magnetic structure would be expected to adopt just one of
those two configurations, which can be directly tested by the
magnetic diffraction data.

The two magnetic reflections observed in the difference
diffraction data in Fig.~\ref{fig:npd} at $d = 3.88$ and 7.67
{\AA} are indexed as (-1,2,0.5) and (0,1,0.5), respectively. The
peak at higher $d$-spacing was found to be significantly broader
than that at $3.88$ {\AA}. We assign this broadening to the
effects of stacking faults, as discussed above. Without a fully
quantitative model of the stacking faults we cannot rule out the
possible existence of otherwise unobserved weak magnetic
reflections close to background levels. However, all statistically
significant reflections can be fit using a peak specific
broadening model, hence allowing for the zigzag and stripy models,
and the moment direction, to be tested.

The solid and dashed lines in Fig.~\ref{fig:npd} show the
calculated diffraction patterns for both magnetic structures. In
both cases the relative intensity of the two observed reflections
could only be reproduced with magnetic moments oriented within the
$ac$-plane, however, within experimental uncertainties the fit to
the data was not sufficiently sensitive to the precise moment
direction in this plane. Furthermore, one can immediately rule out
the stripy model (dashed line), which predicts strong magnetic
reflections for any moment direction at $d$-spacing positions
where no such reflections are observed in the data, beyond any
ambiguity inherent to peak broadening effects. To estimate the
ordered Ru magnetic moment magnitude we calibrate the magnetic
diffraction intensities against the (001) nuclear reflection
(Fig.~\ref{fig:npd} inset), which is unaffected by stacking faults
(see Sec.~\ref{sec:xrd}), and fit the Ru moment magnitude within
the zigzag model (black line in Fig.~\ref{fig:npd}). In the fit to
the reference (001) structural peak only an intensity scale factor
was varied with all the internal crystal structure parameters kept
fixed to the values at 80~K (Table~\ref{struc_tab}), only
adjusting for the effect of the lower temperature in the neutron
measurements by a slight reduction in the lattice parameters,
estimated by fitting the nuclear peak positions observed in the
neutron diffraction data at low $d$-spacing (not shown). Using
this procedure we find the lower limit for the magnetic moment to
be 0.64(4) $\mu_\mathrm{B}$, with the actual value being dependent
on the precise moment direction, which the present data only
constraints to be in the $ac$ plane. Despite not knowing the exact
moment direction, the symmetry of the ground state magnetic
structure is now well established as zigzag in-plane order with
antiferromagnetic stacking along $c$, in qualitative agreement
with previous studies.\cite{sears15,Banerjee15} In monoclinic
symmetry the magnetic structure is described by the magnetic
super-space group $C_c2/m$, with basis transformation
[[1,0,2],[0,-1,0],[0,0,-2]] and origin shift (-1/2,0,-1/2) with
respect to the parent $C2/m$ unit cell.

\subsection{\label{ins}Implications of Monoclinic
Symmetry for the Magnetic Exchange Interactions}

Here we discuss possible implications of the monoclinic crystal
structure for the low-energy spin excitations in the magnetically
ordered phase. Recent inelastic powder neutron scattering
measurements have reported\cite{Banerjee15} dispersive magnetic
excitations above a gap of $\approx$1.7~meV and it was proposed that
features observed in the inelastic spectrum at intermediate
energies above this gap could be understood based on a minimal
Kitaev-Heisenberg model on the honeycomb lattice, with an
antiferromagnetic Kitaev exchange $K$ and a ferromagnetic
Heisenberg term $J$. However, it was pointed out that this minimal
model could not account for the observed spin gap, as for a
honeycomb lattice with full three-fold symmetry (as expected in
the trigonal $P\,3_112$ structural model) the exchanges along the
three bonds meeting at each lattice site are symmetry-equivalent,
and in this case linear spin-wave theory predicts a gapless
spectrum,\cite{Banerjee15} contrary to that observed
experimentally. We note that the $C2/m$ monoclinic structure
breaks the symmetry between the three bonds in the honeycomb
planes, making the $b$-axis bond non-equivalent to the other two
bonds, which remain symmetry-equivalent; this opens the
possibility that the magnitude of the anisotropic exchange could
be different between the two families of bonds. By repeating the
linear spin-wave calculations reported in
Ref.~\onlinecite{Banerjee15} we find that an anisotropy of order
10\% in the magnitude of the Kitaev term between the two families
of bonds (larger in magnitude for the $b$-axis bond) would be
sufficient to account for the magnitude of the observed spin gap,
suggesting that non-equivalence between the different bonds in the
honeycomb plane induced by the underlying monoclinic distortions
may provide a natural mechanism to explain the observed spin gap.

\section{\label{sec:electronic}Electronic structure}
Here we discuss the implications of the monoclinic crystal
structure for the electronic band-structure and the magnetic
ground state of the Ru ions. Within a honeycomb layer the
difference in the atomic positions in the trigonal $P\,3_112$
[Ref.~\onlinecite{stroganov57}] compared to the monoclinic $C2/m$
models is on visual inspection minimal. However, subtleties of the
crystal structure in fact have profound implications for the
nature of the electronic structure. The trigonal crystal structure
features shorter Ru-Ru bonds, and as a result the calculated
electronic structure is dominated by Ru-Ru direct hopping. On the
other hand, in the present monoclinic structure the dominant
hopping process is one $via$ Cl $p$ states, which, as discussed in
references\cite{Mazin2012,Foyevtsova2013,Yingli2015} for
Na$_2$IrO$_3$, leads to the formation of quasimolecular orbitals
(QMO) that consist of a linear combination of $t_{2g}$ states of
the six Ru atoms in a hexagon.

\begin{figure}[htbp]
\includegraphics[width=0.48\textwidth]{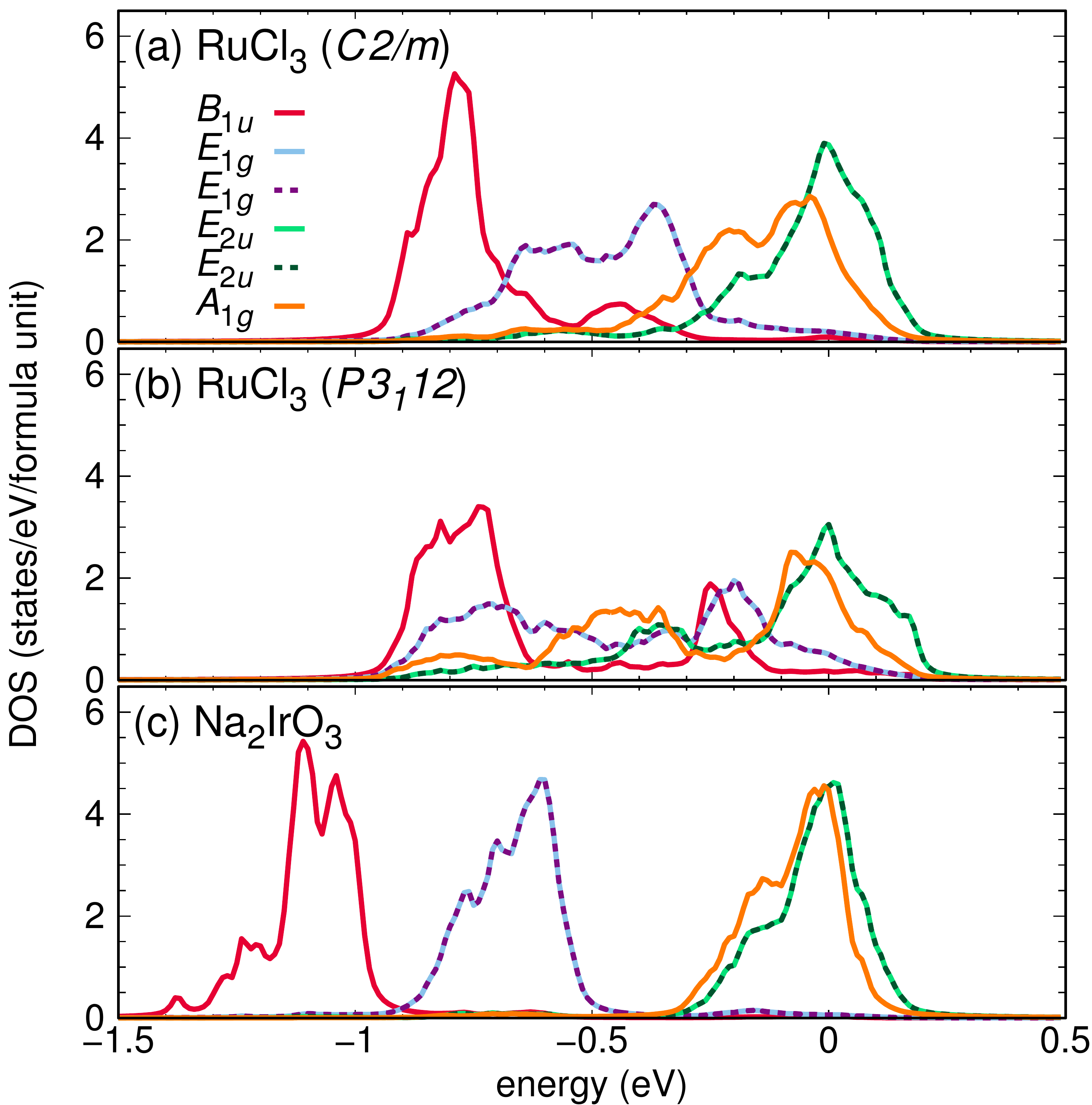}
\caption{(Color online) GGA density of states projected onto the
  quasi-molecular orbital basis of a) $\alpha$-RuCl$_3$ in the $C
  2/m$ structure, b) $\alpha$-RuCl$_3$ in the $P\,3_112$
  structure\cite{stroganov57} and c) Na$_2$IrO$_3$.}
\label{dos_P3112}
\end{figure}

In Fig.~\ref{dos_P3112} we show the {\it nonrelativistic} density
of states within GGA projected onto the QMO basis for
$\alpha$-RuCl$_3$ in the $C 2/m$ and $P\,3_112$ crystal
structures, as well as that for Na$_2$IrO$_3$ for comparison.
While $\alpha$-RuCl$_3$ ($C 2/m$) and Na$_2$IrO$_3$ are
predominantly diagonal in the QMO basis, this is not the case for
$\alpha$-RuCl$_3$ ($P\,3_112$) as can be observed from the strong
mixing of QMO states.

\begin{figure}[htbp]
\includegraphics[width=0.48\textwidth]{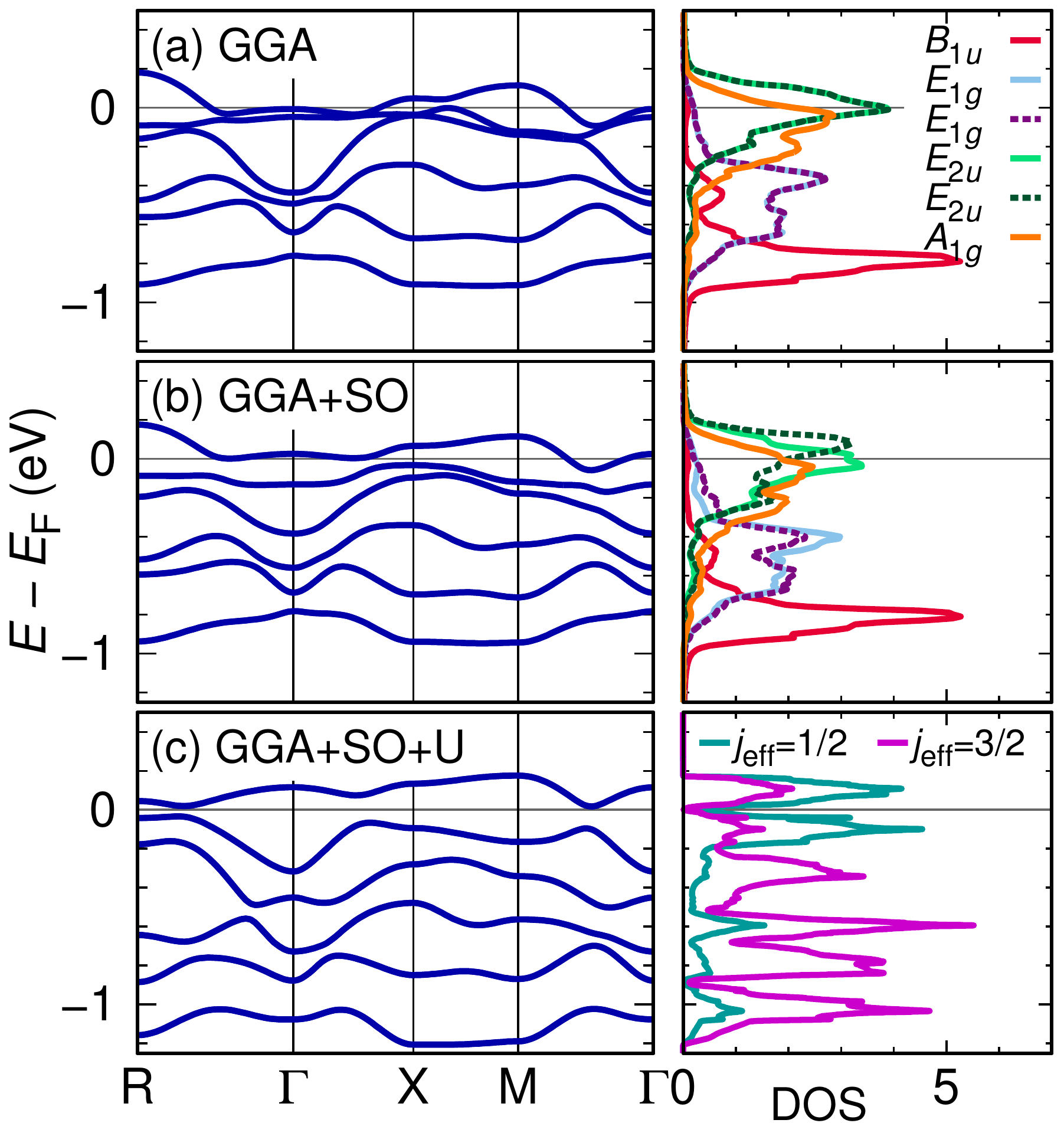}
\caption{(Color online) Band structure and density of states of
$\alpha$-RuCl$_3$ in the $C2/m$ structure obtained within (a) GGA,
(b) GGA+SO, and (c) GGA+SO+U ($U_\mathrm{eff} = 3$~eV). The right
panel shows the projected nonmagnetic GGA and GGA+SO density of
states onto the quasi-molecular orbital basis
\cite{Mazin2012,Foyevtsova2013} and the GGA+SO+U density of states
onto the relativistic $j_\mathrm{eff}$ basis.} \label{band-dos}
\end{figure}

To analyze spin orbit and correlation effects we present in
Fig.~\ref{band-dos} the electronic structure of $\alpha$-RuCl$_3$
($C2/m$) in the GGA, GGA+SO (GGA plus inclusion of spin-orbit
effects) and GGA+SO+U (GGA plus inclusion of spin-orbit effects
and on-site Coulomb repulsion $U$) approximations as implemented
in Wien2k.\cite{Wien2k} Here, an insightful comparison with
Na$_2$IrO$_3$ may be drawn, as follows. In Na$_2$IrO$_3$
[Refs.~\onlinecite{Mazin2012,Foyevtsova2013}] the combination of
accidental degeneracy of the two highest QMOs, $A_{1g}$ and
$E_{2u}$, combined with strong spin-orbit coupling, largely
destroys the QMO and leads instead to the formation of
relativistic $j_\mathrm{eff}=1/2$ orbitals (the QMOs are still
relevant as they generate unexpectedly large second and third
neighbor magnetic interactions\cite{choi12,Chun15}). Adding the
Hubbard $U$ in Na$_2$IrO$_3$ increases the band gap, but does not
affect the electronic structure in any qualitative way. However,
given that the spin-orbit coupling on Ru is much smaller than on
Ir, turning on the spin-orbit coupling leaves the QMO picture in
$\alpha$-RuCl$_3$ ($C 2/m$) almost intact (Fig.~\ref{band-dos}b).
Interestingly, adding $U$ dramatically changes the electronic
structure (Fig.~\ref{band-dos}c). Such an addition effectively
renormalizes the one-electron hopping by a factor of $t/U$ and
increases the effect of spin-orbit coupling that now becomes an
important player. Eventually, the electronic structure with $both$
spin-orbit and $U$ looks surprisingly similar to that of
Na$_2$IrO$_3$
[Refs.~\onlinecite{Mazin2012,Foyevtsova2013,Yingli2015}].

We emphasize that the physics leading to the formation of this
electronic structure in the two systems is qualitatively
different, which needs to be kept in mind when comparing physical
properties of the two compounds. While without spin orbit and
Hubbard correlation both systems are molecular-orbitals solids,
and with inclusion of both effects the spin-orbit interaction
takes control, in Na$_2$IrO$_3$ this happens because the
spin-orbit coupling is initially strong, and correlations play a
secondary role, in $\alpha$-RuCl$_3$ ($C 2/m$) the much stronger
correlation conspires with spin orbit, which otherwise is too weak
to overcome the one-electron hopping effects.

The GGA+SO+U bandstructure for $\alpha$-RuCl$_3$ ($C2/m$) can be
projected onto the $j_\mathrm{eff}$ = 1/2, 3/2 basis as shown in
the density of states in Fig.~\ref{band-dos}c). While there is
some mixing between the two projections, $j_\mathrm{eff}$ = 1/2
has the dominant contribution at the Fermi level. Therefore, a
description of this system in terms of $j_\mathrm{eff}$= 1/2
orbitals may still be valid. This is in qualitative agreement with
GGA+SO+U calculations reported for $\alpha$-RuCl$_3$ in the
$P\,3_112$ structure\cite{Heung15} although the two electronic
structures differ quantitatively.\\

\section{\label{sec:con}Conclusions}

We have proposed a revised three-dimensional crystal structure for
the layered honeycomb magnet \rc based on x-ray diffraction on
un-twinned crystals combined with {\em ab-initio} structural
relaxation calculations. In contrast with the currently-assumed
three-layer stacking periodicity, we have found a single layer
stacking periodicity with a monoclinic unit cell, iso-structural
to \nio, with occasional faults in the stacking sequence. In
powder neutron diffraction and in susceptibility measurements on
both powders and single crystals we have observed a single
magnetic transition near 13~K, and through analysis of the
magnetic diffraction pattern we have confirmed that this phase has
zigzag antiferromagnetic order. Using both static and pulsed
magnetic field experiments we have observed that the zigzag phase
is suppressed by relatively small magnetic fields ($\approx 8$~T)
applied in the honeycomb layers, whereas it is robust in fields
applied perpendicular to the honeycomb layers. We have discussed
how the monoclinic crystal structure could provide a natural
mechanism to explain the spin gap observed in inelastic neutron
scattering experiments and how the asymptotic shape of the
magnetization curve at high fields near saturation is consistent
with proposals for strongly-anisotropic (non-Heisenberg) magnetic
interactions.

\begin{acknowledgements}
We acknowledge useful discussions regarding pulsed field
magnetometry with P.\,A. Goddard, and regarding electronic
structure with G. Khaliullin and S. Winter. Work in Oxford was
supported by EPSRC under Grants No. EP/H014934/1, EP/J003557/1 and
EP/M020517/1, and in Frankfurt by the Deutsche
Forschungsgemeinschaft through grant SFB/TR49. A.A.H. acknowledges support from the Royal Society through an International Newton Fellowship and
Y.L. acknowledges support from a China Scholarship Council (CSC)
Fellowship. I.I.M. was supported by the Office of Naval Research
through the Naval Research Laboratory's Basic Research Program.
R.C. and R.V. were supported in part by KITP under NSF grant
PHY11-25915. Work at LANL was supported by the U.S. Department of
Energy (DoE) Basic Energy Science Field Work Proposal `Science in
100 T'. The NHMFL facility at LANL is funded by the National
Science Foundation Cooperative Agreement No. DMR-1157490, the
State of Florida, and the U.S. DoE. In accordance with the EPSRC
policy framework on research data, access to the data will be made
available from Ref.~\onlinecite{ora}.

\end{acknowledgements}

\bibliography{rucl3_bibliography}

\end{document}